\documentclass[aps,prl,twocolumn,superscriptaddress]{revtex4-1}

\usepackage{mhchem,graphicx,longtable,xfrac}
\usepackage[colorlinks=true,citecolor=blue,linkcolor=blue,breaklinks=true]{hyperref}
\usepackage{amssymb}
\usepackage{graphicx}
\usepackage{amsmath}

\usepackage{times}
\usepackage{color}
\usepackage{bm}

\begin{document}

\newcommand {\beq} {\begin{equation}}
\newcommand {\eeq} {\end{equation}}
\newcommand {\bqa} {\begin{eqnarray}}
\newcommand {\eqa} {\end{eqnarray}}
\newcommand {\ba} {\ensuremath{b^\dagger}}
\newcommand {\Ma} {\ensuremath{M^\dagger}}
\newcommand {\psia} {\ensuremath{\psi^\dagger}}
\newcommand {\psita} {\ensuremath{\tilde{\psi}^\dagger}}
\newcommand{\lp} {\ensuremath{{\lambda '}}}
\newcommand{\A} {\ensuremath{{\bf A}}}
\newcommand{\Q} {\ensuremath{{\bf Q}}}
\newcommand{\kk} {\ensuremath{{\bf k}}}
\newcommand{\qq} {\ensuremath{{\bf q}}}
\newcommand{\kp} {\ensuremath{{\bf k'}}}
\newcommand{\rr} {\ensuremath{{\bf r}}}
\newcommand{\rp} {\ensuremath{{\bf r'}}}
\newcommand {\ep} {\ensuremath{\epsilon}}
\newcommand{\nbr} {\ensuremath{\langle ij \rangle}}
\newcommand {\no} {\nonumber}
\newcommand{\up} {\ensuremath{\uparrow}}
\newcommand{\dn} {\ensuremath{\downarrow}}
\newcommand{\rcol} {\textcolor{red}}


\title{Overdamped antiferromagnetic strange metal state in \ce{Sr_3IrRuO_7}}


\author{Julian L. Schmehr}
\affiliation{Materials Department, University of California, Santa Barbara, California 93106, USA}

\author{Thomas R. Mion}
\affiliation{Department of Physics, Boston College, Chestnut Hill, Massachusetts 02467, USA}

\author{Zach Porter}
\affiliation{Materials Department, University of California, Santa Barbara, California 93106, USA}

\author{Michael Aling}
\affiliation{Materials Department, University of California, Santa Barbara, California 93106, USA}

\author{Huibo Cao}
\affiliation{Neutron Scattering Division, Oak Ridge National Laboratory, Oak Ridge, Tennessee 37831, USA}

\author{Mary H. Upton}
\affiliation{Advanced Photon Source, Argonne National Laboratory, Argonne, Illinois 60439, USA}

\author{Zahirul Islam}
\affiliation{Advanced Photon Source, Argonne National Laboratory, Argonne, Illinois 60439, USA}

\author{Rui-Hua He}
\affiliation{Department of Physics, Boston College, Chestnut Hill, Massachusetts 02467, USA}

\author{Rajdeep Sensarma}
\affiliation{Department of Theoretical Physics, Tata Institute of Fundamental Research, Mumbai 400005, India}

\author{Nandini Trivedi}
\affiliation{Mathematics Department, The Ohio State University, Columbus, OH 43210, USA}

\author{Stephen D. Wilson}
\email[]{stephendwilson@ucsb.edu}
\affiliation{Materials Department, University of California, Santa Barbara, California 93106, USA}


\date{\today}

\begin{abstract}
The unconventional electronic ground state of \ce{Sr_3IrRuO_7} is explored via resonant x-ray scattering techniques and angle-resolved photoemission measurements. As the Ru content approaches $x=0.5$ in \ce{Sr_3(Ir_{1-x}Ru_x)_2O_7}, intermediate to the $J_{eff}=1/2$ Mott state in Sr$_3$Ir$_2$O$_7$ and the quantum critical metal in Sr$_3$Ru$_2$O$_7$, a thermodynamically distinct metallic state emerges.  The electronic structure of this intermediate phase lacks coherent quasiparticles, and charge transport exhibits a linear temperature dependence over a wide range of temperatures. Spin dynamics associated with the long-range antiferromagnetism of this phase show nearly local, overdamped magnetic excitations and an anomalously large energy scale of 200~meV---an energy far in excess of exchange energies present within either the \ce{Sr_3Ir_2O_7} or Sr$_3$Ru$_2$O$_7$ solid-solution endpoints.  Overdamped quasiparticle dynamics driven by strong spin-charge coupling are proposed to explain the incoherent spectral features of the strange metal state in \ce{Sr_3IrRuO_7}.

\end{abstract}


\maketitle

The doping and disorder-induced breakdown of the Mott insulating state in transition metal oxides has long been the subject of experimental and theoretical investigations. This collapse often stabilizes nearby correlated metallic states with unconventional properties, ranging from high temperature superconductivity~\cite{lee2006doping} to pseudogap formation~\cite{timusk1999pseudogap,moreo1999pseudogap,battisti2017universality} to intertwined spin and charge density wave states~\cite{fradkin2015colloquium}. Particularly intriguing are ill-defined ``intermediate" metallic phases that retain the magnetic order of the parent Mott phase after the charge gap has closed, yet prior to the formation of a coherent Fermi liquid. In a Hubbard model, such a state can be induced by filling control~\cite{Yee_2015} and strong disorder also presents an avenue for inducing an intermediate magnetic metallic state~\cite{Heidarian_2004}. In the strong Mott limit, where the charge gap $E_G$ far exceeds the magnetic exchange $J$, this apparent antiferromagnetic metallic phase originates from a local quenching of $E_G$ within a phase separated antiferromagnetic ground state.

Studies of the $J_{eff}=1/2$ Mott states inherent to the Ruddlesden-Popper iridates \ce{Sr_{1+n}Ir_{n}O_{3n+1}} present a new avenue for exploring such intermediate states in the opposite, weak Mott limit, where $E_G\sim J$~\cite{Kim_2008,Kim_2009,Moon_2008}. Here, the charge gap relies on strong, cooperative spin-orbit coupling and crystal electric field effects which allow for a modest on-site Coulomb repulsion $U$ to split the half-filled $J_{eff}=1/2$ valence band. The enhanced covalency inherent to the extended Ir $5d$ valence orbitals of these systems also results in increased hopping, reflected in a strong $J$~\cite{Kim_2012,Moretti_Sala_2015,kim2012magnetic}.  As a result, the bilayer $n=2$ system \ce{Sr_3Ir_2O_7} realizes a Mott state with $E_G/J=1.44$ close to the weak limit $U/W\approx1$~\cite{Kim_2012,Okada_2013}.   

Prior studies exploring the collapse of the Mott state in \ce{Sr_3(Ir_{1-x}Ru_x)_2O_7} via substituting Ir ($5d^5$) with Ru ($4d^4$) cations uncovered an unusual electronic phase diagram~\cite{Dhital_2014}.  The introduction of Ru$^{4+}$ impurities initially generates local, nanoscale, metallic regions which percolate and eventually condense into a \textit{global} metallic phase near $x=0.5$.  AF order endemic to the parent Mott state survives in this global metal, and $T_{AF}$ shows a pronounced maximum near this concentration.  This suggests the formation of an intermediate state where the charge gap has globally collapsed, yet vestiges of the parent Mott state remain.  The origin of this intermediate phase formed between the antiferromagnetic Mott state of \ce{Sr_3Ir_2O_7} and the nonmagnetic Fermi liquid ground state of Sr$_3$Ru$_2$O$_7$~\cite{Ikeda_2000} however remains an open question.  

Here we explore the nature of the AF metallic phase realized within \ce{Sr_3IrRuO_7} by using Ir $L_3$-edge resonant inelastic x-ray scattering (RIXS) and angle-resolved photoemission (ARPES) measurements. ARPES data reveal an incoherent quasiparticle spectrum lacking well-defined quasiparticle peaks, while resonant x-ray scattering data show the persistence of long-range AF order whose dynamics dramatically differ from the AF Mott state of \ce{Sr_3Ir_2O_7}.  Magnons become overdamped, reflecting a strong coupling to the charge carriers, and the disordered metallic state screens long-range spin interactions such that spin excitations form a nearly local band at $194$ meV (substantially larger than the exchange energies of the iridate and ruthenate endpoints). Our data present a unique window into the strong coupling between carrier dynamics and magnetic order realized once the charge gap has collapsed into the intermediate AF metallic phase of a weak Mott state.

 \begin{figure}
 	\includegraphics[width=\columnwidth]{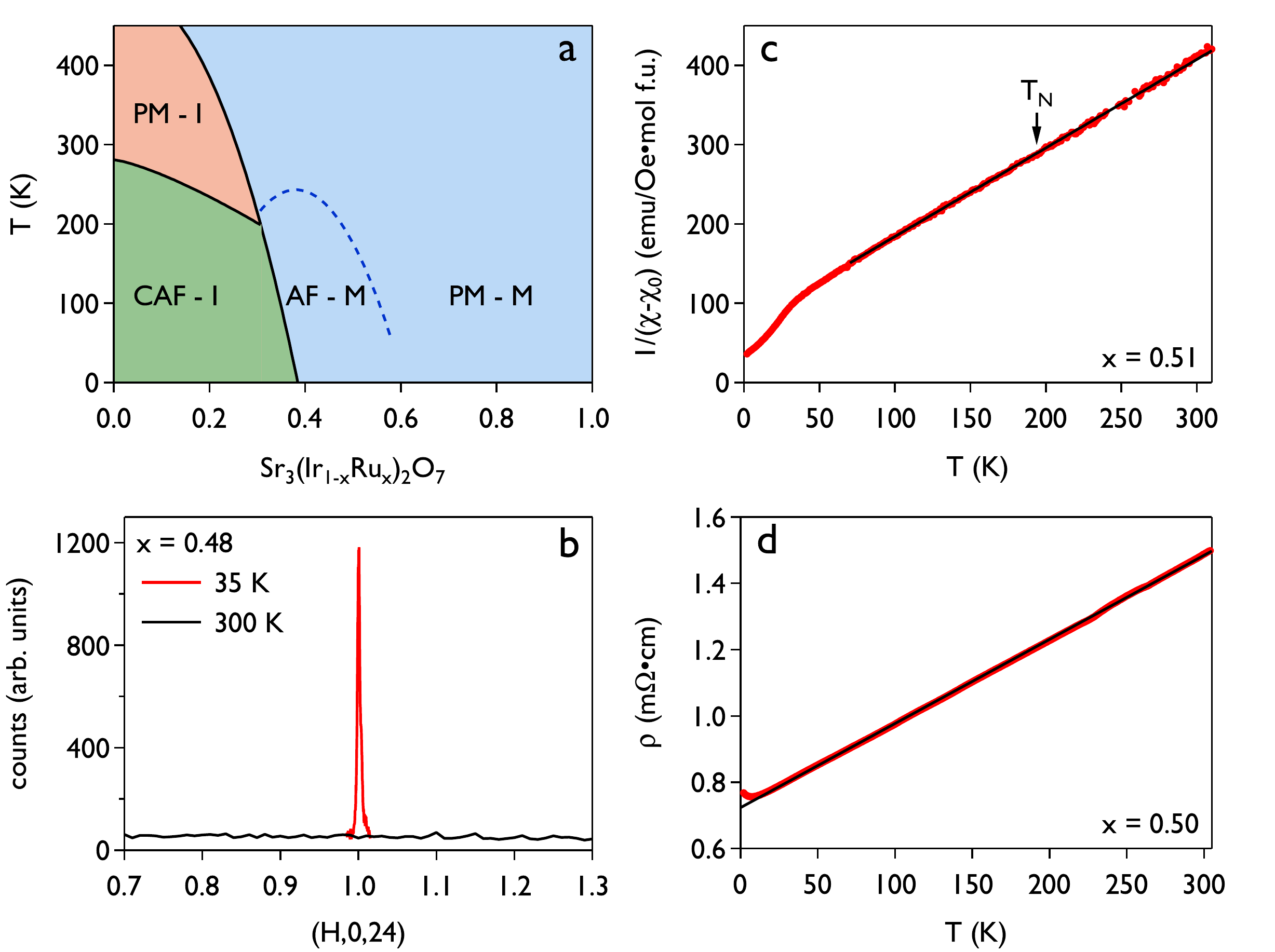}
 	\caption{\label{fig1}(a) Schematic electronic phase diagram of \ce{Sr_3(Ir_{1-x}Ru_x)_2O_7}. (b) Scattering intensity of an $x=0.48$ sample collected at the (1, 0, 24) Bragg position above and below $T_N$. (c) $1/(\chi-\chi_0)$ as a function of temperature for $x=0.51$, with $H_{ab}=100$~Oe. The solid line is the result to a Curie-Weiss fit as detailed in the text. (d) $ab$-plane resistivity plotted as a function of temperature for $x=0.5$.}
 \end{figure}

Crystals of \ce{Sr_3(Ir_{1-x}Ru_x)_2O_7} were grown ~\cite{supplemental,Hogan_2015} and characterized using x-ray diffraction as well as energy dispersive x-ray analysis (EDX). EDX measurements yielded an uncertainty of approximately $1\%$ in Ru composition, and for the purposes of this paper, samples with $x=0.48$, $x=0.5$, and $x=0.51$ are treated as equivalent compositions within the strange metal state. Transport and magnetization data were collected in a Quantum Design PPMS and MPMS3 SQUID magnetometer, respectively. RIXS measurements were performed at the Ir $L_3$ edge ($E=11.215$~keV) on 27-ID-B at the Advanced Photon Source, Argonne National Laboratory~\cite{supplemental}. ARPES data were collected at the Stanford Synchrotron Radiation Lightsource BL 5-4 with linearly polarized 25 eV photons. Neutron scattering experiments were performed on the HB-3A instrument at the High Flux Isotope Reactor, Oak Ridge National Laboratory~\cite{supplemental}.  Momentum space positions are indexed using an orthorhombic $Bbcb$ cell, with lattice parameters $a=5.4820$~\AA, $b=5.4839$~\AA, and $c=20.8962$~\AA.
 
\begin{figure}[t]
	\includegraphics[width=\columnwidth]{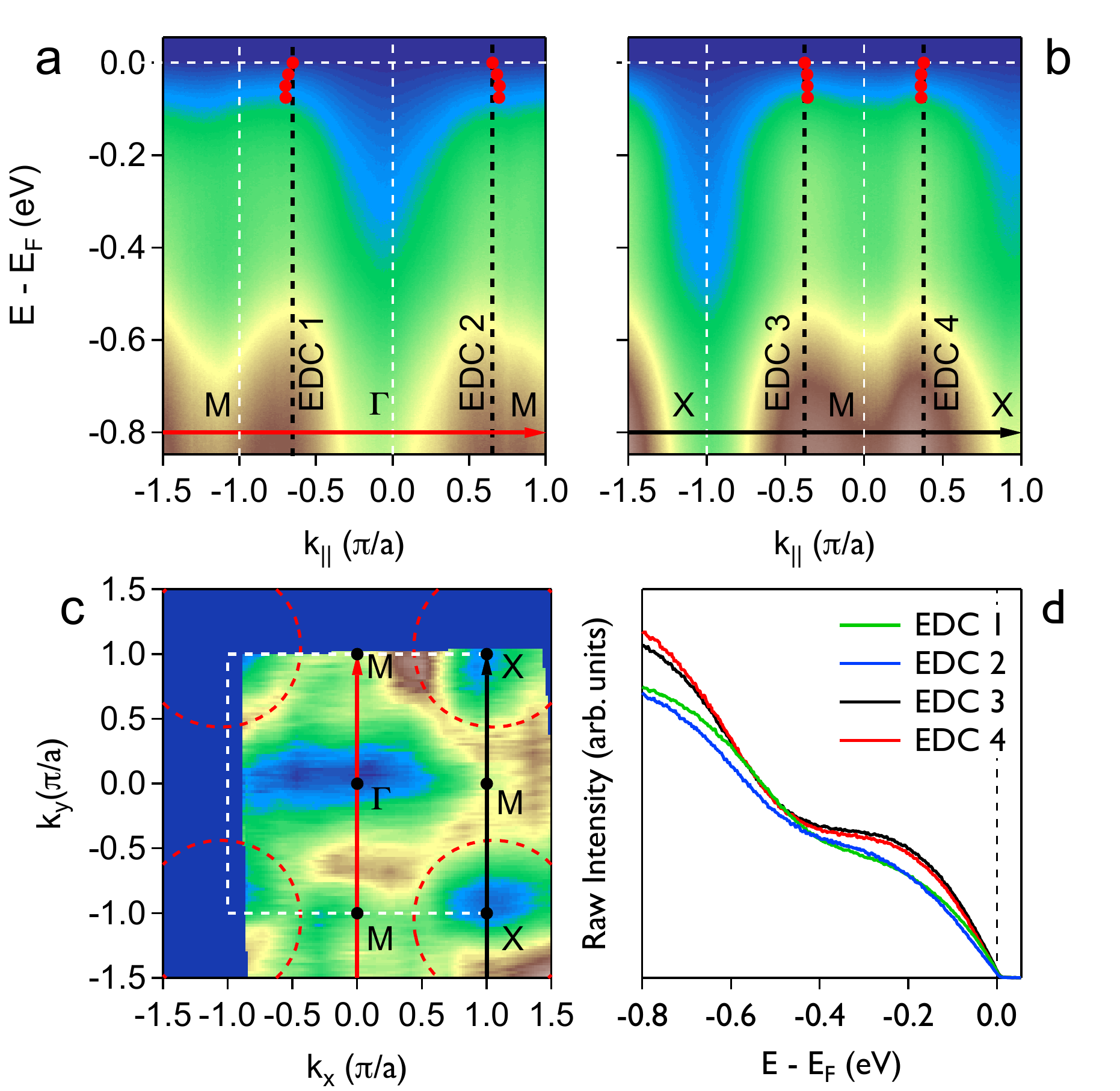}
	\caption{\label{fig2}(a,b) Photoemission spectra collected along high-symmetry directions for $x=0.5$. Red symbols show the dispersions of the hole-like bands near $E-E_F=0$, extracted from peaks in momentum-distribution cuts (see supplementary information~\cite{supplemental}). (c) Map of incoherent spectral weight for $x=0.5$ at $E_F$. Dashed white square shows the Brillouin zone of the $Bbcb$ unit cell, while the high symmetry directions are indicated with respect to the smaller tetragonal unit cell. Red dashed circles highlight features reminiscent of diffuse hole-like pockets centered around the X-points. The arrows indicate the momentum cut directions in (a) and (b). (d) Representative energy distribution curves collected at $k_F$, at positions indicated in figures (a) and (b).}
\end{figure}

The electronic phase diagram of \ce{Sr_3(Ir_{1-x}Ru_x)_2O_7} is reproduced in Fig. \ref{fig1}(a). A globally metallic state with an inhomogeneous local density of states is found near $x=0.5$, and previous neutron scattering results detected the presence of AF correlations in this phase~\cite{Dhital_2014}. Higher resolution, resonant elastic x-ray scattering data plotted in Fig. \ref{fig1}(b) show that this AF state indeed is long-range order.  Momentum scans reveal spin-spin correlation lengths of $\xi_a\approx\xi_c\approx1000$ \AA~ and an isotropic correlation volume~\cite{supplemental}.  This demonstrates that the AF order is a collective property of the metallic state \cite{Dhital_2014}.

Magnetization data plotted in Fig. \ref{fig1}(c) are dominated by the local moments attributable to the Ru-dopants fit to a Curie-Weiss form with a sizable $\Theta_W=-70$~K and $\mu_{eff}=2.68$~$\mu_B$. The size of $\mu_{eff}$ is consistent with the expected concentration 50\% $S=1$ (Ru$^{4+}$) impurities and the absence of local $J_{eff}=1/2$ moments bound by the strong $J$ of this material~\cite{Nagai_2007}.  Notably, there is no signature of Ru moments participating in the onset of AF order near 200 K~\cite{Dhital_2014}. The AF state in the metallic regime is then best envisioned as ordered  $J_{eff}=1/2$ moments within a disordered background of $S=1$ impurities which maintain a local, Curie-Weiss response far below $T_{AF}$.

Electrical resistivity data collected within the $ab$-plane of an $x=0.5$ sample are plotted in Fig. \ref{fig1}(d) and show a linear temperature dependence over a wide range of temperatures above a small, disorder-driven upturn near 10 K. This is emblematic of a strange metal state, and, while the large residual resistivity indicates that disorder-induced scattering is significant in this sample, transport values do not violate the Mott-Ioffe-Regel limit. 

Looking at the metallic state in greater depth, Fig. \ref{fig2}(a,b) shows ARPES spectra collected along high-symmetry directions in an $x=0.5$ sample. These reveal broadened bands dispersing toward the Fermi level ($E_F$). The spectrum consists of seemingly pseudogapped hole pockets centered at the X-points (Fig. \ref{fig2}(c)). Energy distribution curves (EDCs) at the Fermi momenta ($k_F$) along the M-$\Gamma$-M and X-M-X cuts are shown in Fig. \ref{fig2}(d). They possess highly incoherent spectral line shapes with the absence of clear quasiparticle peaks at low energies \cite{supplemental}.

\begin{figure}[t]
	\includegraphics[width=\columnwidth]{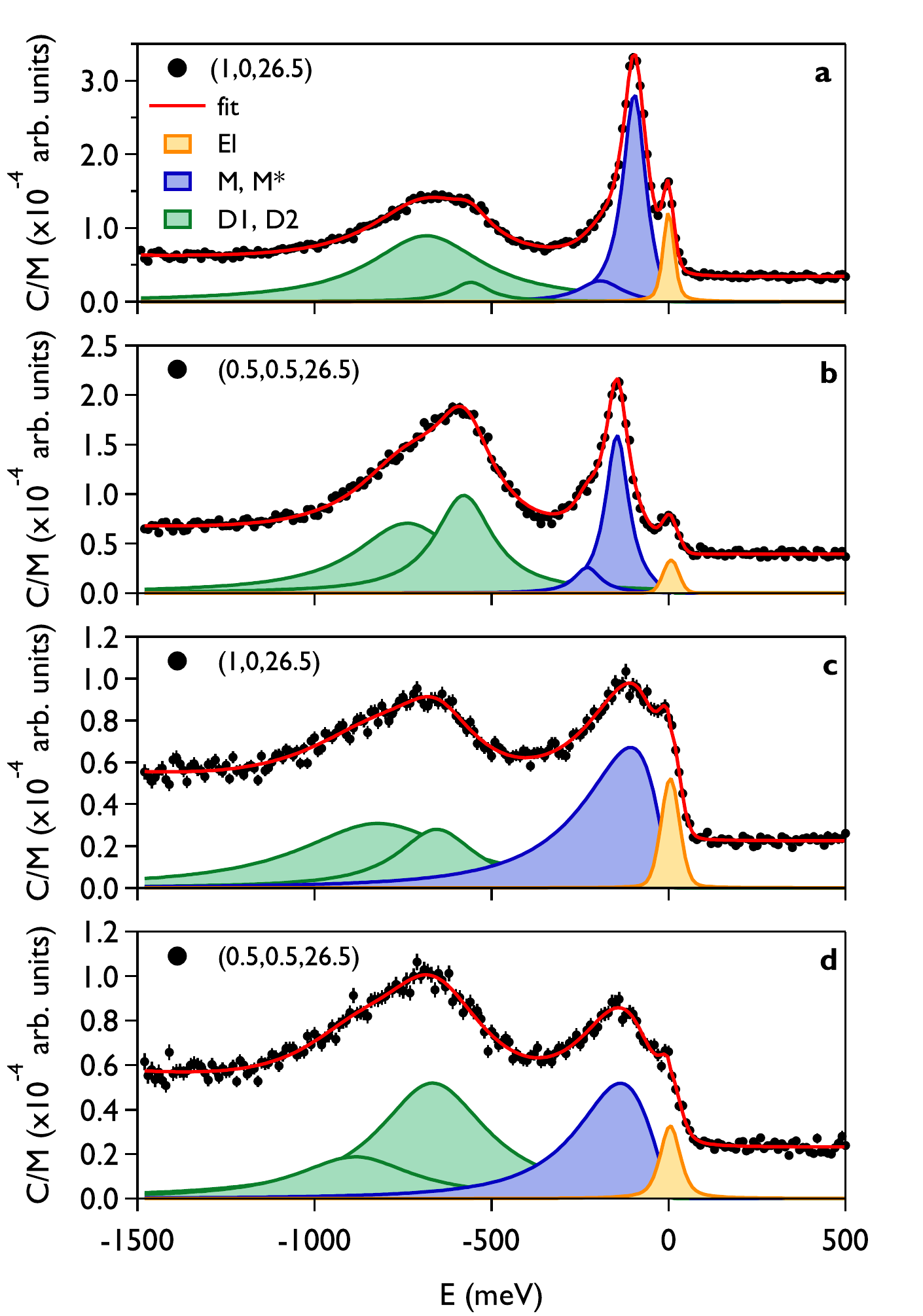}
	\caption{\label{fig3} Representative RIXS spectra for $x=0.17$ sample collected at the magnetic zone center (a) and the zone boundary (b) and for the $x=0.48$ sample at the zone center (c) and boundary (d). All spectra were collected at $T=40$ K.  Solid red line is a fit to the data utilizing the spectral components discussed in the text in addition to a linear background term.}
\end{figure}

While long-range AF order remnant from the $J_{eff}=1/2$ Mott state remains in this metallic phase, the collective excitations underlying the AF state are distinct. Fig. \ref{fig3} shows representative RIXS spectra taken in the ordered phase at the magnetic zone centers and boundaries for an $x=0.48$ crystal.  For comparison, RIXS spectra collected from a concentration within the AF insulating regime ($x=0.17$) are also plotted.

Looking first at Figs. \ref{fig3}(a) and \ref{fig3}(b), the excitations in the AF insulating $x=0.17$ sample are reminiscent of those observed in \ce{Sr_3Ir_2O_7}. An elastic line (El), two high-energy (D1, D2) and two lower-energy features (M, M*) are fit to the data. Consistent with previous reports~\cite{kim2012magnetic, Kim_2012}, these are ascribed to $d\rightarrow d$ crystal field excitations, magnon, and multimagnon excitations respectively. Recent neutron scattering measurements have verified that the M modes sampled by RIXS in the iridates faithfully sample the magnetic excitations \cite{PhysRevB.98.220402}. The M* peak which is conventionally envisioned as a two-magnon feature, is present also in the single layer Sr$_2$IrO$_4$ compound, and is not a second branch of the bilayer system's spin wave dispersion~\cite{kim2012magnetic,PhysRevLett.109.157402}. 

As disorder and carriers induced by Ru substitution may add a damping term to the magnon modes and partially screen extended exchange, the M feature was fit to the form~\cite{Le_Tacon_2011,Monney_2016}:
 \begin{equation}
	\chi''(\omega)=\chi_0'' \frac{\gamma \omega}{(\omega^2-\omega_0^2)^2+\gamma^2\omega^2}.
\end{equation}
Here $\chi_0'' $, $\omega_0$ and $\gamma/2$ are the momentum integrated intensities, characteristic energy scales and damping rates (widths) of the magnetic excitations respectively.

In the AF metallic state, the M feature becomes overdamped with a pronounced high-energy tail. Unlike in the insulating $x=0.17$ crystal where magnons remain in the underdamped regime, the free carriers and disorder in $x=0.48$ induce significant damping within the spin excitations. As a result, for the $x=0.48$ sample, only a single broad M mode is resolved along with the higher energy D1 and D2 modes as shown in Figs. \ref{fig3}(c) and \ref{fig3}(d).  The reduction in Ir-atoms at this substitution level also leads to a strong suppression of the overall intensity. 

\begin{figure}
	\includegraphics[width=\columnwidth]{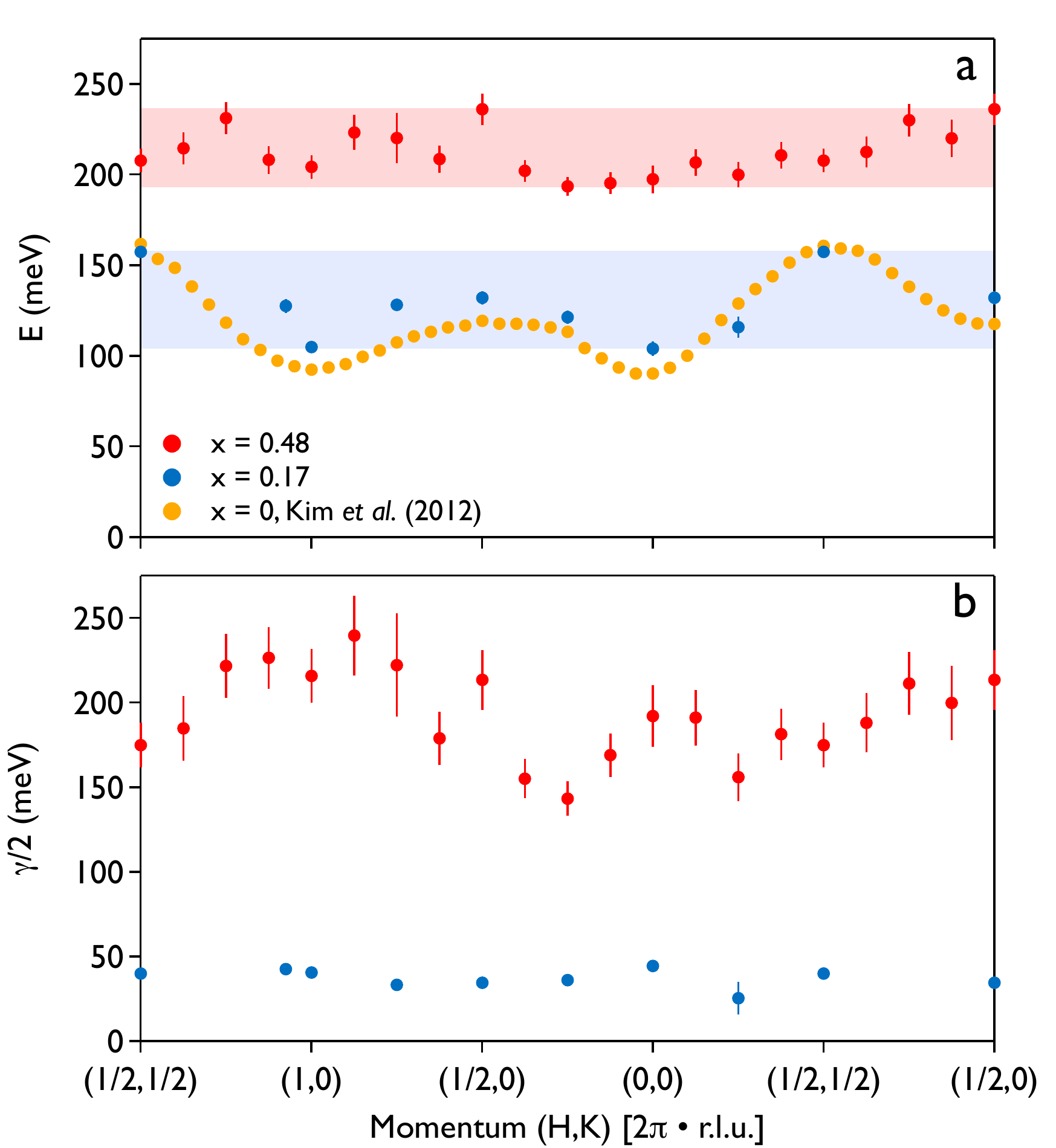}
	\caption{\label{fig4}(a) Characteristic energies $\omega_0$ for the M modes in $x=0$ (reproduced from Ref.~\cite{Kim_2012}), $x=0.17$ and $x=0.48$. Solid line is a fit using an effective mean-field bond operator model as discussed in the text.  (b) Inverse lifetimes $\gamma/2$ of the M modes across the Brillouin zone in both the underdamped $x=0.17$ and overdamped $x=0.48$ samples.}
\end{figure}

The dispersion of the M modes along high symmetry directions is plotted in Fig.  \ref{fig4}(a) for both insulating ($x=0.17$) and metallic samples ($x=0.48$). As a reference, the magnon dispersion of \ce{Sr_3Ir_2O_7} is also reproduced from Ref.~\cite{Kim_2012}. The $x=0.17$ sample shows a dispersion consistent with the parent $x=0$ system, albeit slightly modified via disorder that weakens extended exchange. This renders a slightly enhanced zone center gap value ($E_G=104$ meV) and a reduced magnon bandwidth ($53$ meV). The damping coefficients which are proportional to the inverse lifetimes of the magnon modes are plotted in Fig. \ref{fig4}(b) and are consistent with the spin system being in the underdamped regime ($\gamma/2<\omega_0$).

The dispersion of the $x=0.48$ AF metallic sample is however dramatically renormalized. The excitations become strongly overdamped with a characteristic energy $\omega_0$ pushed \textit{upward} to an AF zone center value of $194$ meV.  Extended exchange interactions are screened, which can be inferred from the flat dispersion along the magnetic zone boundary and the further reduction in bandwidth to 40 meV (a value considerably smaller than $\gamma/2$).  The damping coefficients shown in Fig. \ref{fig4}(b) become comparable to the excitation energy ($\gamma/2 \approx \omega_0$), and the magnons no longer exist as coherent quasiparticles.  We note here that the absence of a resolvable M* feature in the spectra of the $x=0.48$ sample does not appreciably impact the parameterization of its dispersion or the changes in the magnon spectra deduced relative to the $x=0.17$ crystal in the AF insulating phase \cite{supplemental}.

The lattice structure of an $x=0.5$ crystal was investigated via neutron diffraction to rule out any potential changes in bond angles and distances that may impact magnetic exchange. The two endpoints of the phase diagram \ce{Sr_3Ir_2O_7} and Sr$_3$Ru$_2$O$_7$ are structurally similar~\cite{hogan_2016,kiyanagi2004investigation} and interatomic distances, as well as octahedral rotation angles in the $x=0.5$ sample are intermediate between the two endpoints~\cite{supplemental}.  No indication of a preferred site for Ru or Ir was observed, and structural changes are not the underlying cause of the renormalization of magnetic spectrum in \ce{Sr_3IrRuO_7}.  

We next discuss a theoretical model that can explain the following experimental observations in \ce{Sr_3IrRuO_7}: (i) The strong damping of the magnon, with a theoretical estimate of the damping coefficient matching the experimental data, (ii) the non-dispersive peak in the magnon spectral function at a characteristic energy larger than that in the insulating phase, and (iii) the evidence for broad features in the single particle spectral function suggesting the lack of well-defined quasiparticles.

We model \ce{Sr_3IrRuO_7} as a doped Mott insulator with strong spin-orbit coupling. The parent compound possesses $G$-type AF order, with antiparallel spin across bipartite sublattices~\cite{supplemental}. As vacancies move in this magnetic background, the nearest neighbor hopping of electrons across sublattices creates magnetic domain walls that lead to strong inelastic scattering between the vacancies and the collective magnon excitations. The resulting incoherent motion leads to the formation of spin polarons~\cite{martinez1991spin}, consistent with the broad features seen in ARPES. In the presence of spin-orbit coupling, electron spins can flip as it hops across sublattices, so the spin order is less perturbed, but quantum fluctuations invariably generate inelastic processes that broaden the electronic spectral functions.  
 
The strength of the magnon-quasiparticle scattering depends strongly on the ratio between the exchange coupling driving the magnetism and the intersublattice hopping parameter. We estimate the superexchange scale from the bond-order mean-field theory~\cite{Sachdev_bond} used to model the magnon dispersion at half-filling~\cite{Moretti_Sala_2015}. In this case, the largest Heisenberg coupling is along the c-axis with a scale $J_c \sim 90$~meV.  
The largest hopping scale can be obtained from theoretical estimates of a bandwidth ~\cite{Singh_2001} of $\sim 400$~meV in \ce{Sr_3Ru_2O_7}. Alternately, one can use estimates of the Hubbard repulsion $U$ of $1.5-2 eV$ extracted from tunneling gap measurements~\cite{Okada2013} and the superexchange scale obtained earlier to get the largest tunneling parameter $t_c =\sqrt{J_cU}/2 \sim 200$~meV, consistent with a bandwidth of about $400 $meV.

The inelastic scattering of magnons by vacancies is incorporated in the magnon spectral function through a self-energy term. The self-energy~\cite{supplemental} describes the particle-hole susceptibility of vacancies and is much larger than the dispersive magnon features on the energy scale of the superexchange coupling. Hence the magnon spectral function is primarily dictated by the form of the magnon self-energy and not by the magnon dispersion. The incoherent ARPES spectrum leads to non-dispersive features in the magnon self-energy on the scale of the bandwidth $\sim 400$ meV, which is close to the damping scale seen in RIXS experiments. In the itinerant limit, the enhanced Hund's coupling and correlation effects in the strange metal state renormalize the local susceptibility $\chi^{\prime\prime}(\omega)$ into a peaked function (in this case near 200 meV)~\cite{liu2012nature}. The non-dispersive peak in the magnon spectral function is thus related to the incoherent motion of the charges and not to the superexchange interaction between spins.  Comparable damping and coupling of magnons to the charge channel are also reported in doped Mott states of electron-doped high-T$_c$ cuprates, which also have a highly itinerant character~\cite{ishii2014high}.

Enhanced Hund's coupling in the local limit introduced by the addition of $4d$-electron Ru impurities may increase effective onsite and intersite spin-orbit interactions and magnify anisotropic exchange terms as well as gap values~\cite{PhysRevB.90.115122}. \ce{Sr_3(Ir_{1-x}Ru_x)_2O_7} however does not follow the conventional picture of a Hund's metal. Local moment behavior is instead enhanced in \ce{Sr_3(Ir_{1-x}Ru_x)_2O_7} as the system is driven away from half-filling~\cite{Dhital_2014}. 
    
In \ce{Sr_2Ir_{1-x}Ru_xO_4}, Ru substitution leads to a spin-flop transition, coincident with the opening of a sizable gap in the magnetic excitation spectra~\cite{Calder_2015,Calder_2016}. Upon further Ru substitution magnetic order is suppressed completely, while the magnetic excitations become dispersionless with gap values similar to those observed in \ce{Sr_3IrRuO_7}~\cite{Cao_2017}. Strong $S=1$ disorder, together with quenching of extended exchange couplings was invoked to explain the flat band of excitations; however a comparable AF ordered, strange metal state is not reported.  

In summary, the intermediate AF metallic state following the global collapse of the charge gap in \ce{Sr_3(Ir_{1-x}Ru_x)_2O_7} is characterized by strange metal transport with an incoherent quasiparticle spectrum near $E_F$.  Strong coupling between the spin and charge excitations is proposed, which enhances the apparent energy scale of excitations within the spin channel and is the genesis of the unconventional dynamics associated with the AF state in Sr$_3$IrRuO$_7$.   Sr$_3$IrRuO$_7$ therefore stands as an intriguing intermediate link between the global carrier dynamics inherent to the Fermi liquid boundary and the vestigial long-range AF order inherent to the Mott boundary in the \ce{Sr_3(Ir_{1-x}Ru_x)_2O_7} phase diagram.


\begin{acknowledgments}
This work was primarily supported by NSF Award No. DMR-1505549 (S.D.W).  
Additional funding support was provided by ARO Award W911NF-16-1-0361 (J.L.S. and Z.P.) and the National Science Foundation Graduate Research Fellowship Program under Grant No. DGE-1258923 (T.R.M.). 
J.L.S., M. A., and S.D.W. also gratefully acknowledge funding from the W.M. Keck Foundation.  
The MRL Shared Experimental Facilities are supported by the MRSEC Program of the NSF under Award No. DMR 1720256; a member of the NSF-funded Materials Research Facilities Network.  
The work at ORNL's HFIR was sponsored by the Scientific User Facilities Division, Office of Science, Basic Energy Sciences, U.S. Department of Energy. 
This research used resources of the Advanced Photon Source, a U.S. Department of Energy (DOE) Office of Science User Facility operated for the DOE Office of Science by Argonne National Laboratory under Contract No. DE-AC02-06CH11357.

\end{acknowledgments}

\bibliography{bibtexfile}

\begin{thebibliography}{35}%
\makeatletter
\providecommand \@ifxundefined [1]{%
 \@ifx{#1\undefined}
}%
\providecommand \@ifnum [1]{%
 \ifnum #1\expandafter \@firstoftwo
 \else \expandafter \@secondoftwo
 \fi
}%
\providecommand \@ifx [1]{%
 \ifx #1\expandafter \@firstoftwo
 \else \expandafter \@secondoftwo
 \fi
}%
\providecommand \natexlab [1]{#1}%
\providecommand \enquote  [1]{``#1''}%
\providecommand \bibnamefont  [1]{#1}%
\providecommand \bibfnamefont [1]{#1}%
\providecommand \citenamefont [1]{#1}%
\providecommand \href@noop [0]{\@secondoftwo}%
\providecommand \href [0]{\begingroup \@sanitize@url \@href}%
\providecommand \@href[1]{\@@startlink{#1}\@@href}%
\providecommand \@@href[1]{\endgroup#1\@@endlink}%
\providecommand \@sanitize@url [0]{\catcode `\\12\catcode `\$12\catcode
  `\&12\catcode `\#12\catcode `\^12\catcode `\_12\catcode `\%12\relax}%
\providecommand \@@startlink[1]{}%
\providecommand \@@endlink[0]{}%
\providecommand \url  [0]{\begingroup\@sanitize@url \@url }%
\providecommand \@url [1]{\endgroup\@href {#1}{\urlprefix }}%
\providecommand \urlprefix  [0]{URL }%
\providecommand \Eprint [0]{\href }%
\providecommand \doibase [0]{http://dx.doi.org/}%
\providecommand \selectlanguage [0]{\@gobble}%
\providecommand \bibinfo  [0]{\@secondoftwo}%
\providecommand \bibfield  [0]{\@secondoftwo}%
\providecommand \translation [1]{[#1]}%
\providecommand \BibitemOpen [0]{}%
\providecommand \bibitemStop [0]{}%
\providecommand \bibitemNoStop [0]{.\EOS\space}%
\providecommand \EOS [0]{\spacefactor3000\relax}%
\providecommand \BibitemShut  [1]{\csname bibitem#1\endcsname}%
\let\auto@bib@innerbib\@empty
\bibitem [{\citenamefont {Lee}\ \emph {et~al.}(2006)\citenamefont {Lee},
  \citenamefont {Nagaosa},\ and\ \citenamefont {Wen}}]{lee2006doping}%
  \BibitemOpen
  \bibfield  {author} {\bibinfo {author} {\bibfnamefont {P.~A.}\ \bibnamefont
  {Lee}}, \bibinfo {author} {\bibfnamefont {N.}~\bibnamefont {Nagaosa}}, \ and\
  \bibinfo {author} {\bibfnamefont {X.-G.}\ \bibnamefont {Wen}},\ }\href
  {\doibase 10.1103/RevModPhys.78.17} {\bibfield  {journal} {\bibinfo
  {journal} {Rev. Mod. Phys.}\ }\textbf {\bibinfo {volume} {78}},\ \bibinfo
  {pages} {17} (\bibinfo {year} {2006})}\BibitemShut {NoStop}%
\bibitem [{\citenamefont {Timusk}\ and\ \citenamefont
  {Statt}(1999)}]{timusk1999pseudogap}%
  \BibitemOpen
  \bibfield  {author} {\bibinfo {author} {\bibfnamefont {T.}~\bibnamefont
  {Timusk}}\ and\ \bibinfo {author} {\bibfnamefont {B.}~\bibnamefont {Statt}},\
  }\href {http://stacks.iop.org/0034-4885/62/i=1/a=002} {\bibfield  {journal}
  {\bibinfo  {journal} {Rep. Prog. Phys.}\ }\textbf {\bibinfo {volume} {62}},\
  \bibinfo {pages} {61} (\bibinfo {year} {1999})}\BibitemShut {NoStop}%
\bibitem [{\citenamefont {Moreo}\ \emph {et~al.}(1999)\citenamefont {Moreo},
  \citenamefont {Yunoki},\ and\ \citenamefont {Dagotto}}]{moreo1999pseudogap}%
  \BibitemOpen
  \bibfield  {author} {\bibinfo {author} {\bibfnamefont {A.}~\bibnamefont
  {Moreo}}, \bibinfo {author} {\bibfnamefont {S.}~\bibnamefont {Yunoki}}, \
  and\ \bibinfo {author} {\bibfnamefont {E.}~\bibnamefont {Dagotto}},\ }\href
  {\doibase 10.1103/PhysRevLett.83.2773} {\bibfield  {journal} {\bibinfo
  {journal} {Phys. Rev. Lett.}\ }\textbf {\bibinfo {volume} {83}},\ \bibinfo
  {pages} {2773} (\bibinfo {year} {1999})}\BibitemShut {NoStop}%
\bibitem [{\citenamefont {Battisti}\ \emph {et~al.}(2017)\citenamefont
  {Battisti}, \citenamefont {Bastiaans}, \citenamefont {Fedoseev},
  \citenamefont {De~La~Torre}, \citenamefont {Iliopoulos}, \citenamefont
  {Tamai}, \citenamefont {Hunter}, \citenamefont {Perry}, \citenamefont
  {Zaanen}, \citenamefont {Baumberger} \emph
  {et~al.}}]{battisti2017universality}%
  \BibitemOpen
  \bibfield  {author} {\bibinfo {author} {\bibfnamefont {I.}~\bibnamefont
  {Battisti}}, \bibinfo {author} {\bibfnamefont {K.~M.}\ \bibnamefont
  {Bastiaans}}, \bibinfo {author} {\bibfnamefont {V.}~\bibnamefont {Fedoseev}},
  \bibinfo {author} {\bibfnamefont {A.}~\bibnamefont {De~La~Torre}}, \bibinfo
  {author} {\bibfnamefont {N.}~\bibnamefont {Iliopoulos}}, \bibinfo {author}
  {\bibfnamefont {A.}~\bibnamefont {Tamai}}, \bibinfo {author} {\bibfnamefont
  {E.~C.}\ \bibnamefont {Hunter}}, \bibinfo {author} {\bibfnamefont {R.~S.}\
  \bibnamefont {Perry}}, \bibinfo {author} {\bibfnamefont {J.}~\bibnamefont
  {Zaanen}}, \bibinfo {author} {\bibfnamefont {F.}~\bibnamefont {Baumberger}},
  \emph {et~al.},\ }\href {\doibase 10.1038/nphys3894} {\bibfield  {journal}
  {\bibinfo  {journal} {Nat. Phys.}\ }\textbf {\bibinfo {volume} {13}},\
  \bibinfo {pages} {21} (\bibinfo {year} {2017})}\BibitemShut {NoStop}%
\bibitem [{\citenamefont {Fradkin}\ \emph {et~al.}(2015)\citenamefont
  {Fradkin}, \citenamefont {Kivelson},\ and\ \citenamefont
  {Tranquada}}]{fradkin2015colloquium}%
  \BibitemOpen
  \bibfield  {author} {\bibinfo {author} {\bibfnamefont {E.}~\bibnamefont
  {Fradkin}}, \bibinfo {author} {\bibfnamefont {S.~A.}\ \bibnamefont
  {Kivelson}}, \ and\ \bibinfo {author} {\bibfnamefont {J.~M.}\ \bibnamefont
  {Tranquada}},\ }\href {\doibase 10.1103/RevModPhys.87.457} {\bibfield
  {journal} {\bibinfo  {journal} {Rev. Mod. Phys.}\ }\textbf {\bibinfo {volume}
  {87}},\ \bibinfo {pages} {457} (\bibinfo {year} {2015})}\BibitemShut
  {NoStop}%
\bibitem [{\citenamefont {Yee}\ and\ \citenamefont {Balents}(2015)}]{Yee_2015}%
  \BibitemOpen
  \bibfield  {author} {\bibinfo {author} {\bibfnamefont {C.-H.}\ \bibnamefont
  {Yee}}\ and\ \bibinfo {author} {\bibfnamefont {L.}~\bibnamefont {Balents}},\
  }\href {\doibase 10.1103/PhysRevX.5.021007} {\bibfield  {journal} {\bibinfo
  {journal} {Phys. Rev. X}\ }\textbf {\bibinfo {volume} {5}},\ \bibinfo {pages}
  {021007} (\bibinfo {year} {2015})}\BibitemShut {NoStop}%
\bibitem [{\citenamefont {Heidarian}\ and\ \citenamefont
  {Trivedi}(2004)}]{Heidarian_2004}%
  \BibitemOpen
  \bibfield  {author} {\bibinfo {author} {\bibfnamefont {D.}~\bibnamefont
  {Heidarian}}\ and\ \bibinfo {author} {\bibfnamefont {N.}~\bibnamefont
  {Trivedi}},\ }\href {\doibase 10.1103/PhysRevLett.93.126401} {\bibfield
  {journal} {\bibinfo  {journal} {Phys. Rev. Lett.}\ }\textbf {\bibinfo
  {volume} {93}},\ \bibinfo {pages} {126401} (\bibinfo {year}
  {2004})}\BibitemShut {NoStop}%
\bibitem [{\citenamefont {Kim}\ \emph {et~al.}(2008)\citenamefont {Kim},
  \citenamefont {Jin}, \citenamefont {Moon}, \citenamefont {Kim}, \citenamefont
  {Park}, \citenamefont {Leem}, \citenamefont {Yu}, \citenamefont {Noh},
  \citenamefont {Kim}, \citenamefont {Oh}, \citenamefont {Park}, \citenamefont
  {Durairaj}, \citenamefont {Cao},\ and\ \citenamefont {Rotenberg}}]{Kim_2008}%
  \BibitemOpen
  \bibfield  {author} {\bibinfo {author} {\bibfnamefont {B.~J.}\ \bibnamefont
  {Kim}}, \bibinfo {author} {\bibfnamefont {H.}~\bibnamefont {Jin}}, \bibinfo
  {author} {\bibfnamefont {S.~J.}\ \bibnamefont {Moon}}, \bibinfo {author}
  {\bibfnamefont {J.-Y.}\ \bibnamefont {Kim}}, \bibinfo {author} {\bibfnamefont
  {B.-G.}\ \bibnamefont {Park}}, \bibinfo {author} {\bibfnamefont {C.~S.}\
  \bibnamefont {Leem}}, \bibinfo {author} {\bibfnamefont {J.}~\bibnamefont
  {Yu}}, \bibinfo {author} {\bibfnamefont {T.~W.}\ \bibnamefont {Noh}},
  \bibinfo {author} {\bibfnamefont {C.}~\bibnamefont {Kim}}, \bibinfo {author}
  {\bibfnamefont {S.-J.}\ \bibnamefont {Oh}}, \bibinfo {author} {\bibfnamefont
  {J.-H.}\ \bibnamefont {Park}}, \bibinfo {author} {\bibfnamefont
  {V.}~\bibnamefont {Durairaj}}, \bibinfo {author} {\bibfnamefont
  {G.}~\bibnamefont {Cao}}, \ and\ \bibinfo {author} {\bibfnamefont
  {E.}~\bibnamefont {Rotenberg}},\ }\href {\doibase
  10.1103/PhysRevLett.101.076402} {\bibfield  {journal} {\bibinfo  {journal}
  {Phys. Rev. Lett.}\ }\textbf {\bibinfo {volume} {101}},\ \bibinfo {pages}
  {076402} (\bibinfo {year} {2008})}\BibitemShut {NoStop}%
\bibitem [{\citenamefont {Kim}\ \emph {et~al.}(2009)\citenamefont {Kim},
  \citenamefont {Ohsumi}, \citenamefont {Komesu}, \citenamefont {Sakai},
  \citenamefont {Morita}, \citenamefont {Takagi},\ and\ \citenamefont
  {Arima}}]{Kim_2009}%
  \BibitemOpen
  \bibfield  {author} {\bibinfo {author} {\bibfnamefont {B.~J.}\ \bibnamefont
  {Kim}}, \bibinfo {author} {\bibfnamefont {H.}~\bibnamefont {Ohsumi}},
  \bibinfo {author} {\bibfnamefont {T.}~\bibnamefont {Komesu}}, \bibinfo
  {author} {\bibfnamefont {S.}~\bibnamefont {Sakai}}, \bibinfo {author}
  {\bibfnamefont {T.}~\bibnamefont {Morita}}, \bibinfo {author} {\bibfnamefont
  {H.}~\bibnamefont {Takagi}}, \ and\ \bibinfo {author} {\bibfnamefont
  {T.}~\bibnamefont {Arima}},\ }\href {\doibase 10.1126/science.1167106}
  {\bibfield  {journal} {\bibinfo  {journal} {Science}\ }\textbf {\bibinfo
  {volume} {323}},\ \bibinfo {pages} {1329} (\bibinfo {year}
  {2009})}\BibitemShut {NoStop}%
\bibitem [{\citenamefont {Moon}\ \emph {et~al.}(2008)\citenamefont {Moon},
  \citenamefont {Jin}, \citenamefont {Kim}, \citenamefont {Choi}, \citenamefont
  {Lee}, \citenamefont {Yu}, \citenamefont {Cao}, \citenamefont {Sumi},
  \citenamefont {Funakubo}, \citenamefont {Bernhard},\ and\ \citenamefont
  {Noh}}]{Moon_2008}%
  \BibitemOpen
  \bibfield  {author} {\bibinfo {author} {\bibfnamefont {S.~J.}\ \bibnamefont
  {Moon}}, \bibinfo {author} {\bibfnamefont {H.}~\bibnamefont {Jin}}, \bibinfo
  {author} {\bibfnamefont {K.~W.}\ \bibnamefont {Kim}}, \bibinfo {author}
  {\bibfnamefont {W.~S.}\ \bibnamefont {Choi}}, \bibinfo {author}
  {\bibfnamefont {Y.~S.}\ \bibnamefont {Lee}}, \bibinfo {author} {\bibfnamefont
  {J.}~\bibnamefont {Yu}}, \bibinfo {author} {\bibfnamefont {G.}~\bibnamefont
  {Cao}}, \bibinfo {author} {\bibfnamefont {A.}~\bibnamefont {Sumi}}, \bibinfo
  {author} {\bibfnamefont {H.}~\bibnamefont {Funakubo}}, \bibinfo {author}
  {\bibfnamefont {C.}~\bibnamefont {Bernhard}}, \ and\ \bibinfo {author}
  {\bibfnamefont {T.~W.}\ \bibnamefont {Noh}},\ }\href {\doibase
  10.1103/PhysRevLett.101.226402} {\bibfield  {journal} {\bibinfo  {journal}
  {Phys. Rev. Lett.}\ }\textbf {\bibinfo {volume} {101}},\ \bibinfo {pages}
  {226402} (\bibinfo {year} {2008})}\BibitemShut {NoStop}%
\bibitem [{\citenamefont {Kim}\ \emph {et~al.}(2012{\natexlab{a}})\citenamefont
  {Kim}, \citenamefont {Said}, \citenamefont {Casa}, \citenamefont {Upton},
  \citenamefont {Gog}, \citenamefont {Daghofer}, \citenamefont {Jackeli},
  \citenamefont {van~den Brink}, \citenamefont {Khaliullin},\ and\
  \citenamefont {Kim}}]{Kim_2012}%
  \BibitemOpen
  \bibfield  {author} {\bibinfo {author} {\bibfnamefont {J.}~\bibnamefont
  {Kim}}, \bibinfo {author} {\bibfnamefont {A.~H.}\ \bibnamefont {Said}},
  \bibinfo {author} {\bibfnamefont {D.}~\bibnamefont {Casa}}, \bibinfo {author}
  {\bibfnamefont {M.~H.}\ \bibnamefont {Upton}}, \bibinfo {author}
  {\bibfnamefont {T.}~\bibnamefont {Gog}}, \bibinfo {author} {\bibfnamefont
  {M.}~\bibnamefont {Daghofer}}, \bibinfo {author} {\bibfnamefont
  {G.}~\bibnamefont {Jackeli}}, \bibinfo {author} {\bibfnamefont
  {J.}~\bibnamefont {van~den Brink}}, \bibinfo {author} {\bibfnamefont
  {G.}~\bibnamefont {Khaliullin}}, \ and\ \bibinfo {author} {\bibfnamefont
  {B.~J.}\ \bibnamefont {Kim}},\ }\href {\doibase
  10.1103/PhysRevLett.109.157402} {\bibfield  {journal} {\bibinfo  {journal}
  {Phys. Rev. Lett.}\ }\textbf {\bibinfo {volume} {109}},\ \bibinfo {pages}
  {157402} (\bibinfo {year} {2012}{\natexlab{a}})}\BibitemShut {NoStop}%
\bibitem [{\citenamefont {Moretti~Sala}\ \emph {et~al.}(2015)\citenamefont
  {Moretti~Sala}, \citenamefont {Schnells}, \citenamefont {Boseggia},
  \citenamefont {Simonelli}, \citenamefont {Al-Zein}, \citenamefont {Vale},
  \citenamefont {Paolasini}, \citenamefont {Hunter}, \citenamefont {Perry},
  \citenamefont {Prabhakaran}, \citenamefont {Boothroyd}, \citenamefont
  {Krisch}, \citenamefont {Monaco}, \citenamefont {R\o{}nnow}, \citenamefont
  {McMorrow},\ and\ \citenamefont {Mila}}]{Moretti_Sala_2015}%
  \BibitemOpen
  \bibfield  {author} {\bibinfo {author} {\bibfnamefont {M.}~\bibnamefont
  {Moretti~Sala}}, \bibinfo {author} {\bibfnamefont {V.}~\bibnamefont
  {Schnells}}, \bibinfo {author} {\bibfnamefont {S.}~\bibnamefont {Boseggia}},
  \bibinfo {author} {\bibfnamefont {L.}~\bibnamefont {Simonelli}}, \bibinfo
  {author} {\bibfnamefont {A.}~\bibnamefont {Al-Zein}}, \bibinfo {author}
  {\bibfnamefont {J.~G.}\ \bibnamefont {Vale}}, \bibinfo {author}
  {\bibfnamefont {L.}~\bibnamefont {Paolasini}}, \bibinfo {author}
  {\bibfnamefont {E.~C.}\ \bibnamefont {Hunter}}, \bibinfo {author}
  {\bibfnamefont {R.~S.}\ \bibnamefont {Perry}}, \bibinfo {author}
  {\bibfnamefont {D.}~\bibnamefont {Prabhakaran}}, \bibinfo {author}
  {\bibfnamefont {A.~T.}\ \bibnamefont {Boothroyd}}, \bibinfo {author}
  {\bibfnamefont {M.}~\bibnamefont {Krisch}}, \bibinfo {author} {\bibfnamefont
  {G.}~\bibnamefont {Monaco}}, \bibinfo {author} {\bibfnamefont {H.~M.}\
  \bibnamefont {R\o{}nnow}}, \bibinfo {author} {\bibfnamefont {D.~F.}\
  \bibnamefont {McMorrow}}, \ and\ \bibinfo {author} {\bibfnamefont
  {F.}~\bibnamefont {Mila}},\ }\href {\doibase 10.1103/PhysRevB.92.024405}
  {\bibfield  {journal} {\bibinfo  {journal} {Phys. Rev. B}\ }\textbf {\bibinfo
  {volume} {92}},\ \bibinfo {pages} {024405} (\bibinfo {year}
  {2015})}\BibitemShut {NoStop}%
\bibitem [{\citenamefont {Kim}\ \emph {et~al.}(2012{\natexlab{b}})\citenamefont
  {Kim}, \citenamefont {Casa}, \citenamefont {Upton}, \citenamefont {Gog},
  \citenamefont {Kim}, \citenamefont {Mitchell}, \citenamefont {van
  Veenendaal}, \citenamefont {Daghofer}, \citenamefont {van~den Brink},
  \citenamefont {Khaliullin},\ and\ \citenamefont {Kim}}]{kim2012magnetic}%
  \BibitemOpen
  \bibfield  {author} {\bibinfo {author} {\bibfnamefont {J.}~\bibnamefont
  {Kim}}, \bibinfo {author} {\bibfnamefont {D.}~\bibnamefont {Casa}}, \bibinfo
  {author} {\bibfnamefont {M.~H.}\ \bibnamefont {Upton}}, \bibinfo {author}
  {\bibfnamefont {T.}~\bibnamefont {Gog}}, \bibinfo {author} {\bibfnamefont
  {Y.-J.}\ \bibnamefont {Kim}}, \bibinfo {author} {\bibfnamefont {J.~F.}\
  \bibnamefont {Mitchell}}, \bibinfo {author} {\bibfnamefont {M.}~\bibnamefont
  {van Veenendaal}}, \bibinfo {author} {\bibfnamefont {M.}~\bibnamefont
  {Daghofer}}, \bibinfo {author} {\bibfnamefont {J.}~\bibnamefont {van~den
  Brink}}, \bibinfo {author} {\bibfnamefont {G.}~\bibnamefont {Khaliullin}}, \
  and\ \bibinfo {author} {\bibfnamefont {B.~J.}\ \bibnamefont {Kim}},\ }\href
  {\doibase 10.1103/PhysRevLett.108.177003} {\bibfield  {journal} {\bibinfo
  {journal} {Phys. Rev. Lett.}\ }\textbf {\bibinfo {volume} {108}},\ \bibinfo
  {pages} {177003} (\bibinfo {year} {2012}{\natexlab{b}})}\BibitemShut
  {NoStop}%
\bibitem [{\citenamefont {Okada}\ \emph {et~al.}(2000)\citenamefont {Okada},
  \citenamefont {Walkup}, \citenamefont {Lin}, \citenamefont {Dhital},
  \citenamefont {Chang}, \citenamefont {Khadka}, \citenamefont {Zhou},
  \citenamefont {Jeng}, \citenamefont {Paranjape}, \citenamefont {Bansil},
  \citenamefont {Wang}, \citenamefont {Wilson},\ and\ \citenamefont
  {Madhavan}}]{Okada_2013}%
  \BibitemOpen
  \bibfield  {author} {\bibinfo {author} {\bibfnamefont {Y.}~\bibnamefont
  {Okada}}, \bibinfo {author} {\bibfnamefont {D.}~\bibnamefont {Walkup}},
  \bibinfo {author} {\bibfnamefont {H.}~\bibnamefont {Lin}}, \bibinfo {author}
  {\bibfnamefont {C.}~\bibnamefont {Dhital}}, \bibinfo {author} {\bibfnamefont
  {T.-R.}\ \bibnamefont {Chang}}, \bibinfo {author} {\bibfnamefont
  {S.}~\bibnamefont {Khadka}}, \bibinfo {author} {\bibfnamefont
  {W.}~\bibnamefont {Zhou}}, \bibinfo {author} {\bibfnamefont {H.-T.}\
  \bibnamefont {Jeng}}, \bibinfo {author} {\bibfnamefont {M.}~\bibnamefont
  {Paranjape}}, \bibinfo {author} {\bibfnamefont {A.}~\bibnamefont {Bansil}},
  \bibinfo {author} {\bibfnamefont {Z.}~\bibnamefont {Wang}}, \bibinfo {author}
  {\bibfnamefont {S.}~\bibnamefont {Wilson}}, \ and\ \bibinfo {author}
  {\bibfnamefont {V.}~\bibnamefont {Madhavan}},\ }\href {\doibase
  10.1103/PhysRevB.62.R6089} {\bibfield  {journal} {\bibinfo  {journal} {Phys.
  Rev. B}\ }\textbf {\bibinfo {volume} {62}},\ \bibinfo {pages} {R6089}
  (\bibinfo {year} {2000})}\BibitemShut {NoStop}%
\bibitem [{\citenamefont {Dhital}\ \emph {et~al.}(2014)\citenamefont {Dhital},
  \citenamefont {Hogan}, \citenamefont {Zhou}, \citenamefont {Chen},
  \citenamefont {Ren}, \citenamefont {Pokharel}, \citenamefont {Okada},
  \citenamefont {Heine}, \citenamefont {Tian}, \citenamefont {Yamani},
  \citenamefont {Opeil}, \citenamefont {Helton}, \citenamefont {Lynn},
  \citenamefont {Wang}, \citenamefont {Madhavan},\ and\ \citenamefont
  {Wilson}}]{Dhital_2014}%
  \BibitemOpen
  \bibfield  {author} {\bibinfo {author} {\bibfnamefont {C.}~\bibnamefont
  {Dhital}}, \bibinfo {author} {\bibfnamefont {T.}~\bibnamefont {Hogan}},
  \bibinfo {author} {\bibfnamefont {W.}~\bibnamefont {Zhou}}, \bibinfo {author}
  {\bibfnamefont {X.}~\bibnamefont {Chen}}, \bibinfo {author} {\bibfnamefont
  {Z.}~\bibnamefont {Ren}}, \bibinfo {author} {\bibfnamefont {M.}~\bibnamefont
  {Pokharel}}, \bibinfo {author} {\bibfnamefont {Y.}~\bibnamefont {Okada}},
  \bibinfo {author} {\bibfnamefont {M.}~\bibnamefont {Heine}}, \bibinfo
  {author} {\bibfnamefont {W.}~\bibnamefont {Tian}}, \bibinfo {author}
  {\bibfnamefont {Z.}~\bibnamefont {Yamani}}, \bibinfo {author} {\bibfnamefont
  {C.}~\bibnamefont {Opeil}}, \bibinfo {author} {\bibfnamefont {J.~S.}\
  \bibnamefont {Helton}}, \bibinfo {author} {\bibfnamefont {J.~W.}\
  \bibnamefont {Lynn}}, \bibinfo {author} {\bibfnamefont {Z.}~\bibnamefont
  {Wang}}, \bibinfo {author} {\bibfnamefont {V.}~\bibnamefont {Madhavan}}, \
  and\ \bibinfo {author} {\bibfnamefont {S.~D.}\ \bibnamefont {Wilson}},\
  }\href {\doibase 10.1038/ncomms4377} {\bibfield  {journal} {\bibinfo
  {journal} {Nat. Commun.}\ }\textbf {\bibinfo {volume} {5}},\ \bibinfo {pages}
  {3377} (\bibinfo {year} {2014})}\BibitemShut {NoStop}%
\bibitem [{\citenamefont {Ikeda}\ \emph {et~al.}(2013)\citenamefont {Ikeda},
  \citenamefont {Maeno}, \citenamefont {Nakatsuji}, \citenamefont {Kosaka},\
  and\ \citenamefont {Uwatoko}}]{Ikeda_2000}%
  \BibitemOpen
  \bibfield  {author} {\bibinfo {author} {\bibfnamefont {S.-I.}\ \bibnamefont
  {Ikeda}}, \bibinfo {author} {\bibfnamefont {Y.}~\bibnamefont {Maeno}},
  \bibinfo {author} {\bibfnamefont {S.}~\bibnamefont {Nakatsuji}}, \bibinfo
  {author} {\bibfnamefont {M.}~\bibnamefont {Kosaka}}, \ and\ \bibinfo {author}
  {\bibfnamefont {Y.}~\bibnamefont {Uwatoko}},\ }\href
  {https://www.nature.com/articles/nmat3653} {\bibfield  {journal} {\bibinfo
  {journal} {‎Nat. Mater}\ }\textbf {\bibinfo {volume} {12}},\ \bibinfo
  {pages} {707} (\bibinfo {year} {2013})}\BibitemShut {NoStop}%
\bibitem [{sup()}]{supplemental}%
  \BibitemOpen
  \href@noop {} {}\bibinfo {note} {See supplemental information}\BibitemShut
  {NoStop}%
\bibitem [{\citenamefont {Hogan}\ \emph {et~al.}(2015)\citenamefont {Hogan},
  \citenamefont {Yamani}, \citenamefont {Walkup}, \citenamefont {Chen},
  \citenamefont {Dally}, \citenamefont {Ward}, \citenamefont {Dean},
  \citenamefont {Hill}, \citenamefont {Islam}, \citenamefont {Madhavan},\ and\
  \citenamefont {Wilson}}]{Hogan_2015}%
  \BibitemOpen
  \bibfield  {author} {\bibinfo {author} {\bibfnamefont {T.}~\bibnamefont
  {Hogan}}, \bibinfo {author} {\bibfnamefont {Z.}~\bibnamefont {Yamani}},
  \bibinfo {author} {\bibfnamefont {D.}~\bibnamefont {Walkup}}, \bibinfo
  {author} {\bibfnamefont {X.}~\bibnamefont {Chen}}, \bibinfo {author}
  {\bibfnamefont {R.}~\bibnamefont {Dally}}, \bibinfo {author} {\bibfnamefont
  {T.~Z.}\ \bibnamefont {Ward}}, \bibinfo {author} {\bibfnamefont {M.~P.~M.}\
  \bibnamefont {Dean}}, \bibinfo {author} {\bibfnamefont {J.}~\bibnamefont
  {Hill}}, \bibinfo {author} {\bibfnamefont {Z.}~\bibnamefont {Islam}},
  \bibinfo {author} {\bibfnamefont {V.}~\bibnamefont {Madhavan}}, \ and\
  \bibinfo {author} {\bibfnamefont {S.~D.}\ \bibnamefont {Wilson}},\ }\href
  {\doibase 10.1103/PhysRevLett.114.257203} {\bibfield  {journal} {\bibinfo
  {journal} {Phys. Rev. Lett.}\ }\textbf {\bibinfo {volume} {114}},\ \bibinfo
  {pages} {257203} (\bibinfo {year} {2015})}\BibitemShut {NoStop}%
\bibitem [{\citenamefont {Nagai}\ \emph {et~al.}(2007)\citenamefont {Nagai},
  \citenamefont {Yoshida}, \citenamefont {Ikeda}, \citenamefont {Matsuhata},
  \citenamefont {Kito},\ and\ \citenamefont {Kosaka}}]{Nagai_2007}%
  \BibitemOpen
  \bibfield  {author} {\bibinfo {author} {\bibfnamefont {I.}~\bibnamefont
  {Nagai}}, \bibinfo {author} {\bibfnamefont {Y.}~\bibnamefont {Yoshida}},
  \bibinfo {author} {\bibfnamefont {S.~I.}\ \bibnamefont {Ikeda}}, \bibinfo
  {author} {\bibfnamefont {H.}~\bibnamefont {Matsuhata}}, \bibinfo {author}
  {\bibfnamefont {H.}~\bibnamefont {Kito}}, \ and\ \bibinfo {author}
  {\bibfnamefont {M.}~\bibnamefont {Kosaka}},\ }\href
  {http://stacks.iop.org/0953-8984/19/i=13/a=136214} {\bibfield  {journal}
  {\bibinfo  {journal} {J. Phys.: Condens. Matter}\ }\textbf {\bibinfo {volume}
  {19}},\ \bibinfo {pages} {136214} (\bibinfo {year} {2007})}\BibitemShut
  {NoStop}%
\bibitem [{\citenamefont {Calder}\ \emph {et~al.}(2018)\citenamefont {Calder},
  \citenamefont {Pajerowski}, \citenamefont {Stone},\ and\ \citenamefont
  {May}}]{PhysRevB.98.220402}%
  \BibitemOpen
  \bibfield  {author} {\bibinfo {author} {\bibfnamefont {S.}~\bibnamefont
  {Calder}}, \bibinfo {author} {\bibfnamefont {D.~M.}\ \bibnamefont
  {Pajerowski}}, \bibinfo {author} {\bibfnamefont {M.~B.}\ \bibnamefont
  {Stone}}, \ and\ \bibinfo {author} {\bibfnamefont {A.~F.}\ \bibnamefont
  {May}},\ }\href {\doibase 10.1103/PhysRevB.98.220402} {\bibfield  {journal}
  {\bibinfo  {journal} {Phys. Rev. B}\ }\textbf {\bibinfo {volume} {98}},\
  \bibinfo {pages} {220402} (\bibinfo {year} {2018})}\BibitemShut {NoStop}%
\bibitem [{\citenamefont {Kim}\ \emph {et~al.}(2012{\natexlab{c}})\citenamefont
  {Kim}, \citenamefont {Said}, \citenamefont {Casa}, \citenamefont {Upton},
  \citenamefont {Gog}, \citenamefont {Daghofer}, \citenamefont {Jackeli},
  \citenamefont {van~den Brink}, \citenamefont {Khaliullin},\ and\
  \citenamefont {Kim}}]{PhysRevLett.109.157402}%
  \BibitemOpen
  \bibfield  {author} {\bibinfo {author} {\bibfnamefont {J.}~\bibnamefont
  {Kim}}, \bibinfo {author} {\bibfnamefont {A.~H.}\ \bibnamefont {Said}},
  \bibinfo {author} {\bibfnamefont {D.}~\bibnamefont {Casa}}, \bibinfo {author}
  {\bibfnamefont {M.~H.}\ \bibnamefont {Upton}}, \bibinfo {author}
  {\bibfnamefont {T.}~\bibnamefont {Gog}}, \bibinfo {author} {\bibfnamefont
  {M.}~\bibnamefont {Daghofer}}, \bibinfo {author} {\bibfnamefont
  {G.}~\bibnamefont {Jackeli}}, \bibinfo {author} {\bibfnamefont
  {J.}~\bibnamefont {van~den Brink}}, \bibinfo {author} {\bibfnamefont
  {G.}~\bibnamefont {Khaliullin}}, \ and\ \bibinfo {author} {\bibfnamefont
  {B.~J.}\ \bibnamefont {Kim}},\ }\href {\doibase
  10.1103/PhysRevLett.109.157402} {\bibfield  {journal} {\bibinfo  {journal}
  {Phys. Rev. Lett.}\ }\textbf {\bibinfo {volume} {109}},\ \bibinfo {pages}
  {157402} (\bibinfo {year} {2012}{\natexlab{c}})}\BibitemShut {NoStop}%
\bibitem [{\citenamefont {Le~Tacon}\ \emph {et~al.}(2011)\citenamefont
  {Le~Tacon}, \citenamefont {Ghiringhelli}, \citenamefont {Chaloupka},
  \citenamefont {Sala}, \citenamefont {Hinkov}, \citenamefont {Haverkort},
  \citenamefont {Minola}, \citenamefont {Bakr}, \citenamefont {Zhou},
  \citenamefont {Blanco-Canosa}, \citenamefont {Monney}, \citenamefont {Song},
  \citenamefont {Sun}, \citenamefont {Lin}, \citenamefont {De~Luca},
  \citenamefont {Salluzzo}, \citenamefont {Khaliullin}, \citenamefont
  {Schmitt}, \citenamefont {Braicovich},\ and\ \citenamefont
  {Keimer}}]{Le_Tacon_2011}%
  \BibitemOpen
  \bibfield  {author} {\bibinfo {author} {\bibfnamefont {M.}~\bibnamefont
  {Le~Tacon}}, \bibinfo {author} {\bibfnamefont {G.}~\bibnamefont
  {Ghiringhelli}}, \bibinfo {author} {\bibfnamefont {J.}~\bibnamefont
  {Chaloupka}}, \bibinfo {author} {\bibfnamefont {M.~M.}\ \bibnamefont {Sala}},
  \bibinfo {author} {\bibfnamefont {V.}~\bibnamefont {Hinkov}}, \bibinfo
  {author} {\bibfnamefont {M.~W.}\ \bibnamefont {Haverkort}}, \bibinfo {author}
  {\bibfnamefont {M.}~\bibnamefont {Minola}}, \bibinfo {author} {\bibfnamefont
  {M.}~\bibnamefont {Bakr}}, \bibinfo {author} {\bibfnamefont {K.~J.}\
  \bibnamefont {Zhou}}, \bibinfo {author} {\bibfnamefont {S.}~\bibnamefont
  {Blanco-Canosa}}, \bibinfo {author} {\bibfnamefont {C.}~\bibnamefont
  {Monney}}, \bibinfo {author} {\bibfnamefont {Y.~T.}\ \bibnamefont {Song}},
  \bibinfo {author} {\bibfnamefont {G.~L.}\ \bibnamefont {Sun}}, \bibinfo
  {author} {\bibfnamefont {C.~T.}\ \bibnamefont {Lin}}, \bibinfo {author}
  {\bibfnamefont {G.~M.}\ \bibnamefont {De~Luca}}, \bibinfo {author}
  {\bibfnamefont {M.}~\bibnamefont {Salluzzo}}, \bibinfo {author}
  {\bibfnamefont {G.}~\bibnamefont {Khaliullin}}, \bibinfo {author}
  {\bibfnamefont {T.}~\bibnamefont {Schmitt}}, \bibinfo {author} {\bibfnamefont
  {L.}~\bibnamefont {Braicovich}}, \ and\ \bibinfo {author} {\bibfnamefont
  {B.}~\bibnamefont {Keimer}},\ }\href {http://dx.doi.org/10.1038/nphys2041}
  {\bibfield  {journal} {\bibinfo  {journal} {Nat. Phys.}\ }\textbf {\bibinfo
  {volume} {7}},\ \bibinfo {pages} {725 EP } (\bibinfo {year}
  {2011})}\BibitemShut {NoStop}%
\bibitem [{\citenamefont {Monney}\ \emph {et~al.}(2016)\citenamefont {Monney},
  \citenamefont {Schmitt}, \citenamefont {Matt}, \citenamefont {Mesot},
  \citenamefont {Strocov}, \citenamefont {Lipscombe}, \citenamefont {Hayden},\
  and\ \citenamefont {Chang}}]{Monney_2016}%
  \BibitemOpen
  \bibfield  {author} {\bibinfo {author} {\bibfnamefont {C.}~\bibnamefont
  {Monney}}, \bibinfo {author} {\bibfnamefont {T.}~\bibnamefont {Schmitt}},
  \bibinfo {author} {\bibfnamefont {C.~E.}\ \bibnamefont {Matt}}, \bibinfo
  {author} {\bibfnamefont {J.}~\bibnamefont {Mesot}}, \bibinfo {author}
  {\bibfnamefont {V.~N.}\ \bibnamefont {Strocov}}, \bibinfo {author}
  {\bibfnamefont {O.~J.}\ \bibnamefont {Lipscombe}}, \bibinfo {author}
  {\bibfnamefont {S.~M.}\ \bibnamefont {Hayden}}, \ and\ \bibinfo {author}
  {\bibfnamefont {J.}~\bibnamefont {Chang}},\ }\href {\doibase
  10.1103/PhysRevB.93.075103} {\bibfield  {journal} {\bibinfo  {journal} {Phys.
  Rev. B}\ }\textbf {\bibinfo {volume} {93}},\ \bibinfo {pages} {075103}
  (\bibinfo {year} {2016})}\BibitemShut {NoStop}%
\bibitem [{\citenamefont {Hogan}\ \emph {et~al.}(2016)\citenamefont {Hogan},
  \citenamefont {Bjaalie}, \citenamefont {Zhao}, \citenamefont {Belvin},
  \citenamefont {Wang}, \citenamefont {Van~de Walle}, \citenamefont {Hsieh},\
  and\ \citenamefont {Wilson}}]{hogan_2016}%
  \BibitemOpen
  \bibfield  {author} {\bibinfo {author} {\bibfnamefont {T.}~\bibnamefont
  {Hogan}}, \bibinfo {author} {\bibfnamefont {L.}~\bibnamefont {Bjaalie}},
  \bibinfo {author} {\bibfnamefont {L.}~\bibnamefont {Zhao}}, \bibinfo {author}
  {\bibfnamefont {C.}~\bibnamefont {Belvin}}, \bibinfo {author} {\bibfnamefont
  {X.}~\bibnamefont {Wang}}, \bibinfo {author} {\bibfnamefont {C.~G.}\
  \bibnamefont {Van~de Walle}}, \bibinfo {author} {\bibfnamefont
  {D.}~\bibnamefont {Hsieh}}, \ and\ \bibinfo {author} {\bibfnamefont {S.~D.}\
  \bibnamefont {Wilson}},\ }\href {\doibase 10.1103/PhysRevB.93.134110}
  {\bibfield  {journal} {\bibinfo  {journal} {Phys. Rev. B}\ }\textbf {\bibinfo
  {volume} {93}},\ \bibinfo {pages} {134110} (\bibinfo {year}
  {2016})}\BibitemShut {NoStop}%
\bibitem [{\citenamefont {Kiyanagi}\ \emph {et~al.}(2004)\citenamefont
  {Kiyanagi}, \citenamefont {Tsuda}, \citenamefont {Aso}, \citenamefont
  {Kimura}, \citenamefont {Noda}, \citenamefont {Yoshida}, \citenamefont
  {Ikeda},\ and\ \citenamefont {Uwatoko}}]{kiyanagi2004investigation}%
  \BibitemOpen
  \bibfield  {author} {\bibinfo {author} {\bibfnamefont {R.}~\bibnamefont
  {Kiyanagi}}, \bibinfo {author} {\bibfnamefont {K.}~\bibnamefont {Tsuda}},
  \bibinfo {author} {\bibfnamefont {N.}~\bibnamefont {Aso}}, \bibinfo {author}
  {\bibfnamefont {H.}~\bibnamefont {Kimura}}, \bibinfo {author} {\bibfnamefont
  {Y.}~\bibnamefont {Noda}}, \bibinfo {author} {\bibfnamefont {Y.}~\bibnamefont
  {Yoshida}}, \bibinfo {author} {\bibfnamefont {S.-I.}\ \bibnamefont {Ikeda}},
  \ and\ \bibinfo {author} {\bibfnamefont {Y.}~\bibnamefont {Uwatoko}},\ }\href
  {https://doi.org/10.1143/JPSJ.73.639} {\bibfield  {journal} {\bibinfo
  {journal} {J. Phys. Soc. Jpn.}\ }\textbf {\bibinfo {volume} {73}},\ \bibinfo
  {pages} {639} (\bibinfo {year} {2004})}\BibitemShut {NoStop}%
\bibitem [{\citenamefont {Martinez}\ and\ \citenamefont
  {Horsch}(1991)}]{martinez1991spin}%
  \BibitemOpen
  \bibfield  {author} {\bibinfo {author} {\bibfnamefont {G.}~\bibnamefont
  {Martinez}}\ and\ \bibinfo {author} {\bibfnamefont {P.}~\bibnamefont
  {Horsch}},\ }\href@noop {} {\bibfield  {journal} {\bibinfo  {journal}
  {Physical Review B}\ }\textbf {\bibinfo {volume} {44}},\ \bibinfo {pages}
  {317} (\bibinfo {year} {1991})}\BibitemShut {NoStop}%
\bibitem [{\citenamefont {Sachdev}\ and\ \citenamefont
  {Bhatt}(1990)}]{Sachdev_bond}%
  \BibitemOpen
  \bibfield  {author} {\bibinfo {author} {\bibfnamefont {S.}~\bibnamefont
  {Sachdev}}\ and\ \bibinfo {author} {\bibfnamefont {R.~N.}\ \bibnamefont
  {Bhatt}},\ }\href {\doibase 10.1103/PhysRevB.41.9323} {\bibfield  {journal}
  {\bibinfo  {journal} {Phys. Rev. B}\ }\textbf {\bibinfo {volume} {41}},\
  \bibinfo {pages} {9323} (\bibinfo {year} {1990})}\BibitemShut {NoStop}%
\bibitem [{\citenamefont {Singh}\ and\ \citenamefont
  {Mazin}(2001)}]{Singh_2001}%
  \BibitemOpen
  \bibfield  {author} {\bibinfo {author} {\bibfnamefont {D.~J.}\ \bibnamefont
  {Singh}}\ and\ \bibinfo {author} {\bibfnamefont {I.~I.}\ \bibnamefont
  {Mazin}},\ }\href {\doibase 10.1103/PhysRevB.63.165101} {\bibfield  {journal}
  {\bibinfo  {journal} {Phys. Rev. B}\ }\textbf {\bibinfo {volume} {63}},\
  \bibinfo {pages} {165101} (\bibinfo {year} {2001})}\BibitemShut {NoStop}%
\bibitem [{\citenamefont {Okada}\ \emph {et~al.}(2013)\citenamefont {Okada},
  \citenamefont {Walkup}, \citenamefont {Lin}, \citenamefont {Dhital},
  \citenamefont {Chang}, \citenamefont {Khadka}, \citenamefont {Zhou},
  \citenamefont {Jeng}, \citenamefont {Paranjape}, \citenamefont {Bansil},
  \citenamefont {Wang}, \citenamefont {Wilson},\ and\ \citenamefont
  {Madhavan}}]{Okada2013}%
  \BibitemOpen
  \bibfield  {author} {\bibinfo {author} {\bibfnamefont {Y.}~\bibnamefont
  {Okada}}, \bibinfo {author} {\bibfnamefont {D.}~\bibnamefont {Walkup}},
  \bibinfo {author} {\bibfnamefont {H.}~\bibnamefont {Lin}}, \bibinfo {author}
  {\bibfnamefont {C.}~\bibnamefont {Dhital}}, \bibinfo {author} {\bibfnamefont
  {T.-R.}\ \bibnamefont {Chang}}, \bibinfo {author} {\bibfnamefont
  {S.}~\bibnamefont {Khadka}}, \bibinfo {author} {\bibfnamefont
  {W.}~\bibnamefont {Zhou}}, \bibinfo {author} {\bibfnamefont {H.-T.}\
  \bibnamefont {Jeng}}, \bibinfo {author} {\bibfnamefont {M.}~\bibnamefont
  {Paranjape}}, \bibinfo {author} {\bibfnamefont {A.}~\bibnamefont {Bansil}},
  \bibinfo {author} {\bibfnamefont {Z.}~\bibnamefont {Wang}}, \bibinfo {author}
  {\bibfnamefont {S.~D.}\ \bibnamefont {Wilson}}, \ and\ \bibinfo {author}
  {\bibfnamefont {V.}~\bibnamefont {Madhavan}},\ }\href@noop {} {\bibfield
  {journal} {\bibinfo  {journal} {Nature Materials}\ }\textbf {\bibinfo
  {volume} {12}},\ \bibinfo {pages} {707} (\bibinfo {year} {2013})}\BibitemShut
  {NoStop}%
\bibitem [{\citenamefont {Liu}\ \emph {et~al.}(2012)\citenamefont {Liu},
  \citenamefont {Harriger}, \citenamefont {Luo}, \citenamefont {Wang},
  \citenamefont {Ewings}, \citenamefont {Guidi}, \citenamefont {Park},
  \citenamefont {Haule}, \citenamefont {Kotliar}, \citenamefont {Hayden} \emph
  {et~al.}}]{liu2012nature}%
  \BibitemOpen
  \bibfield  {author} {\bibinfo {author} {\bibfnamefont {M.}~\bibnamefont
  {Liu}}, \bibinfo {author} {\bibfnamefont {L.~W.}\ \bibnamefont {Harriger}},
  \bibinfo {author} {\bibfnamefont {H.}~\bibnamefont {Luo}}, \bibinfo {author}
  {\bibfnamefont {M.}~\bibnamefont {Wang}}, \bibinfo {author} {\bibfnamefont
  {R.}~\bibnamefont {Ewings}}, \bibinfo {author} {\bibfnamefont
  {T.}~\bibnamefont {Guidi}}, \bibinfo {author} {\bibfnamefont
  {H.}~\bibnamefont {Park}}, \bibinfo {author} {\bibfnamefont {K.}~\bibnamefont
  {Haule}}, \bibinfo {author} {\bibfnamefont {G.}~\bibnamefont {Kotliar}},
  \bibinfo {author} {\bibfnamefont {S.}~\bibnamefont {Hayden}},  \emph
  {et~al.},\ }\href@noop {} {\bibfield  {journal} {\bibinfo  {journal} {Nature
  Physics}\ }\textbf {\bibinfo {volume} {8}},\ \bibinfo {pages} {376} (\bibinfo
  {year} {2012})}\BibitemShut {NoStop}%
\bibitem [{\citenamefont {Ishii}\ \emph {et~al.}(2014)\citenamefont {Ishii},
  \citenamefont {Fujita}, \citenamefont {Sasaki}, \citenamefont {Minola},
  \citenamefont {Dellea}, \citenamefont {Mazzoli}, \citenamefont {Kummer},
  \citenamefont {Ghiringhelli}, \citenamefont {Braicovich}, \citenamefont
  {Tohyama} \emph {et~al.}}]{ishii2014high}%
  \BibitemOpen
  \bibfield  {author} {\bibinfo {author} {\bibfnamefont {K.}~\bibnamefont
  {Ishii}}, \bibinfo {author} {\bibfnamefont {M.}~\bibnamefont {Fujita}},
  \bibinfo {author} {\bibfnamefont {T.}~\bibnamefont {Sasaki}}, \bibinfo
  {author} {\bibfnamefont {M.}~\bibnamefont {Minola}}, \bibinfo {author}
  {\bibfnamefont {G.}~\bibnamefont {Dellea}}, \bibinfo {author} {\bibfnamefont
  {C.}~\bibnamefont {Mazzoli}}, \bibinfo {author} {\bibfnamefont
  {K.}~\bibnamefont {Kummer}}, \bibinfo {author} {\bibfnamefont
  {G.}~\bibnamefont {Ghiringhelli}}, \bibinfo {author} {\bibfnamefont
  {L.}~\bibnamefont {Braicovich}}, \bibinfo {author} {\bibfnamefont
  {T.}~\bibnamefont {Tohyama}},  \emph {et~al.},\ }\href {\doibase
  10.1038/ncomms4714} {\bibfield  {journal} {\bibinfo  {journal} {Nat.
  Commun.}\ }\textbf {\bibinfo {volume} {5}},\ \bibinfo {pages} {3714}
  (\bibinfo {year} {2014})}\BibitemShut {NoStop}%
\bibitem [{\citenamefont {Isobe}\ and\ \citenamefont
  {Nagaosa}(2014)}]{PhysRevB.90.115122}%
  \BibitemOpen
  \bibfield  {author} {\bibinfo {author} {\bibfnamefont {H.}~\bibnamefont
  {Isobe}}\ and\ \bibinfo {author} {\bibfnamefont {N.}~\bibnamefont
  {Nagaosa}},\ }\href {\doibase 10.1103/PhysRevB.90.115122} {\bibfield
  {journal} {\bibinfo  {journal} {Phys. Rev. B}\ }\textbf {\bibinfo {volume}
  {90}},\ \bibinfo {pages} {115122} (\bibinfo {year} {2014})}\BibitemShut
  {NoStop}%
\bibitem [{\citenamefont {Calder}\ \emph {et~al.}(2015)\citenamefont {Calder},
  \citenamefont {Kim}, \citenamefont {Cao}, \citenamefont {Cantoni},
  \citenamefont {May}, \citenamefont {Cao}, \citenamefont {Aczel},
  \citenamefont {Matsuda}, \citenamefont {Choi}, \citenamefont {Haskel},
  \citenamefont {Sales}, \citenamefont {Mandrus}, \citenamefont {Lumsden},\
  and\ \citenamefont {Christianson}}]{Calder_2015}%
  \BibitemOpen
  \bibfield  {author} {\bibinfo {author} {\bibfnamefont {S.}~\bibnamefont
  {Calder}}, \bibinfo {author} {\bibfnamefont {J.~W.}\ \bibnamefont {Kim}},
  \bibinfo {author} {\bibfnamefont {G.-X.}\ \bibnamefont {Cao}}, \bibinfo
  {author} {\bibfnamefont {C.}~\bibnamefont {Cantoni}}, \bibinfo {author}
  {\bibfnamefont {A.~F.}\ \bibnamefont {May}}, \bibinfo {author} {\bibfnamefont
  {H.~B.}\ \bibnamefont {Cao}}, \bibinfo {author} {\bibfnamefont {A.~A.}\
  \bibnamefont {Aczel}}, \bibinfo {author} {\bibfnamefont {M.}~\bibnamefont
  {Matsuda}}, \bibinfo {author} {\bibfnamefont {Y.}~\bibnamefont {Choi}},
  \bibinfo {author} {\bibfnamefont {D.}~\bibnamefont {Haskel}}, \bibinfo
  {author} {\bibfnamefont {B.~C.}\ \bibnamefont {Sales}}, \bibinfo {author}
  {\bibfnamefont {D.}~\bibnamefont {Mandrus}}, \bibinfo {author} {\bibfnamefont
  {M.~D.}\ \bibnamefont {Lumsden}}, \ and\ \bibinfo {author} {\bibfnamefont
  {A.~D.}\ \bibnamefont {Christianson}},\ }\href {\doibase
  10.1103/PhysRevB.92.165128} {\bibfield  {journal} {\bibinfo  {journal} {Phys.
  Rev. B}\ }\textbf {\bibinfo {volume} {92}},\ \bibinfo {pages} {165128}
  (\bibinfo {year} {2015})}\BibitemShut {NoStop}%
\bibitem [{\citenamefont {Calder}\ \emph {et~al.}(2016)\citenamefont {Calder},
  \citenamefont {Kim}, \citenamefont {Taylor}, \citenamefont {Upton},
  \citenamefont {Casa}, \citenamefont {Cao}, \citenamefont {Mandrus},
  \citenamefont {Lumsden},\ and\ \citenamefont {Christianson}}]{Calder_2016}%
  \BibitemOpen
  \bibfield  {author} {\bibinfo {author} {\bibfnamefont {S.}~\bibnamefont
  {Calder}}, \bibinfo {author} {\bibfnamefont {J.~W.}\ \bibnamefont {Kim}},
  \bibinfo {author} {\bibfnamefont {A.~E.}\ \bibnamefont {Taylor}}, \bibinfo
  {author} {\bibfnamefont {M.~H.}\ \bibnamefont {Upton}}, \bibinfo {author}
  {\bibfnamefont {D.}~\bibnamefont {Casa}}, \bibinfo {author} {\bibfnamefont
  {G.}~\bibnamefont {Cao}}, \bibinfo {author} {\bibfnamefont {D.}~\bibnamefont
  {Mandrus}}, \bibinfo {author} {\bibfnamefont {M.~D.}\ \bibnamefont
  {Lumsden}}, \ and\ \bibinfo {author} {\bibfnamefont {A.~D.}\ \bibnamefont
  {Christianson}},\ }\href {\doibase 10.1103/PhysRevB.94.220407} {\bibfield
  {journal} {\bibinfo  {journal} {Phys. Rev. B}\ }\textbf {\bibinfo {volume}
  {94}},\ \bibinfo {pages} {220407} (\bibinfo {year} {2016})}\BibitemShut
  {NoStop}%
\bibitem [{\citenamefont {Cao}\ \emph {et~al.}(2017)\citenamefont {Cao},
  \citenamefont {Liu}, \citenamefont {Xu}, \citenamefont {Yin}, \citenamefont
  {Meyers}, \citenamefont {Kim}, \citenamefont {Casa}, \citenamefont {Upton},
  \citenamefont {Gog}, \citenamefont {Berlijn}, \citenamefont {Alvarez},
  \citenamefont {Yuan}, \citenamefont {Terzic}, \citenamefont {Tranquada},
  \citenamefont {Hill}, \citenamefont {Cao}, \citenamefont {Konik},\ and\
  \citenamefont {Dean}}]{Cao_2017}%
  \BibitemOpen
  \bibfield  {author} {\bibinfo {author} {\bibfnamefont {Y.}~\bibnamefont
  {Cao}}, \bibinfo {author} {\bibfnamefont {X.}~\bibnamefont {Liu}}, \bibinfo
  {author} {\bibfnamefont {W.}~\bibnamefont {Xu}}, \bibinfo {author}
  {\bibfnamefont {W.-G.}\ \bibnamefont {Yin}}, \bibinfo {author} {\bibfnamefont
  {D.}~\bibnamefont {Meyers}}, \bibinfo {author} {\bibfnamefont
  {J.}~\bibnamefont {Kim}}, \bibinfo {author} {\bibfnamefont {D.}~\bibnamefont
  {Casa}}, \bibinfo {author} {\bibfnamefont {M.~H.}\ \bibnamefont {Upton}},
  \bibinfo {author} {\bibfnamefont {T.}~\bibnamefont {Gog}}, \bibinfo {author}
  {\bibfnamefont {T.}~\bibnamefont {Berlijn}}, \bibinfo {author} {\bibfnamefont
  {G.}~\bibnamefont {Alvarez}}, \bibinfo {author} {\bibfnamefont
  {S.}~\bibnamefont {Yuan}}, \bibinfo {author} {\bibfnamefont {J.}~\bibnamefont
  {Terzic}}, \bibinfo {author} {\bibfnamefont {J.~M.}\ \bibnamefont
  {Tranquada}}, \bibinfo {author} {\bibfnamefont {J.~P.}\ \bibnamefont {Hill}},
  \bibinfo {author} {\bibfnamefont {G.}~\bibnamefont {Cao}}, \bibinfo {author}
  {\bibfnamefont {R.~M.}\ \bibnamefont {Konik}}, \ and\ \bibinfo {author}
  {\bibfnamefont {M.~P.~M.}\ \bibnamefont {Dean}},\ }\href {\doibase
  10.1103/PhysRevB.95.121103} {\bibfield  {journal} {\bibinfo  {journal} {Phys.
  Rev. B}\ }\textbf {\bibinfo {volume} {95}},\ \bibinfo {pages} {121103}
  (\bibinfo {year} {2017})}\BibitemShut {NoStop}%
\end{thebibliography}%

\end{document}